\documentclass[
    ,final            % use final for the camera ready runs
  ]
  {aipproc}

\layoutstyle{8x11double}

\begin{document}

\title{Nonperturbative anharmonic phenomena in crystal lattice dynamics}

\author{M. I. Katsnelson}{
  address={Department of Physics, Uppsala University, Box 530, SE-751 21 Uppsala,
  Sweden}
}

\author{A. V. Trefilov}{
  address={Russian Science Center ``Kurchatov Institute'', 123182, Moscow, Russia}
}

\begin{abstract}
 Slow dynamics of energy transfer between different phonon modes under the
 resonance conditions is considered. It may result in new effects in the
 inelastic and quasielastic neutron scattering spectra.
\end{abstract}

\maketitle

%%%%%%%%%%%%%%%%%%%%%%%%%%%%%%%%%%%%%%%%%%%%
%% MAINMATTER
%%%%%%%%%%%%%%%%%%%%%%%%%%%%%%%%%%%%%%%%%%%%

\section{Introduction}

Crystal lattice dynamics \cite{BH,KT} is a prototype theory for many-body
physics as a whole. In this case all principal approximations and concepts
can be justified accurately on the basis of adiabatic (Born-Oppenheimer)
approximation of quantum mechanics. Due to the smallness of typical atomic
displacements in crystals $\overline{u}$ in comparison with interatomic
distances $d$ one can pass rigorously from a problem of strongly interacting 
\textit{particles} (atoms, ions, or molecules) to a problem of \textit{weakly%
} interacting \textit{quasiparticles} (phonons). In the leading order in the
smallness parameter $\eta =\overline{u}/d$ the crystal lattice dynamics and
thermodynamics can be described in terms of an \textit{ideal }phonon gas
(harmonic approximation) \cite{BH,KT}. With the temperature $T$ increase the
parameter $\eta $ increases as well, however, due to a semiempirical
Lindemann criterion (see, e.g., \cite{KT,brat} and Refs. therein) $\eta
\simeq 0.1$ at the melting point $T=T_{m}$ so higher-order (\textit{%
anharmonic}) contributions to thermodynamic properties are usually small up
to the melting temperature \cite{VKTtherm,KMT}. This statement is true for
the most of \textit{average} characteristics. At the same time, for some 
\textit{peculiar} modes anharmonicity can be crucially strong, especially in
the vicinity of some structural transformations (see, e.g., computational
results for Ba \cite{Ba} and Zr \cite{Zr1,Zr2}, as well as inelastic neutron
scattering data for A15 structure compounds \cite{A15} and for
high-temperature bcc phases of Ti, Zr, and Hf \cite{Petry}). For these cases
a standard phonon picture is not adequate, for instance, additional
non-phonon peaks in the dynamical structural factor may appear \cite{gorn96}%
. A ``slow'' (low-frequency) lattice dynamics in the form of a ``central
peak'' in quasielastic neutron scattering is also typical for these strongly
anharmonic modes \cite{A15,AxeZrNb,plakida}. 

Phonon picture can be broken for these modes since the corresponding
effective potential $V\left( u\right) $ turns out to be essentially
nonparabolic \cite{Ba,Zr1,gorn96,plakida,FMM87}. In the vicinity of
structural phase transitions (e.g., ferroelectric transitions \cite{plakida}
or martensitic transformations in metals \cite{Ba,FMM87}) this potential
normally has several minima corresponding to several competing phases, the
height of the barrier being much smaller than a typical cohesion energy. It
is not surprising that the harmonic approximation is an extremally poor
starting point to describe this situation. Here we discuss the following
issue: is the phonon picture always adequate in a \textit{generic} case when 
$\eta <<1$ for all phonon modes? To answer this question it is instructive
to consider a crystal from the point of view of a contemporary theory of
dynamic systems rather than from that of statistical thermodynamics \cite
{FMM87}. According to the KAM (Kolmogorov-Arnold-Moser) theory \cite
{KAM1} a system of noninteracting oscillators (that is, an ideal
phonon gas) is stable with respect to small interphonon interactions
(anharmonicity) except the cases of the resonance where the ratio of phonon
frequencies is close to the ratio of small enough integer numbers. One can
expect that this exceptional case is a situation where even small
anharmonicities can lead to essentially nonperturbative effects in the
lattice dynamics \cite{FMM87,JETPLett87,JETP90}. 

\section{Resonance phenomena in phonon subsystem}

The well-known Fermi resonance in the Raman molecular spectra under the
conditions of integer ratio of phonon frequencies \cite{Fermi,KT} is
a prototype example of these nonperturbative resonance phenomena. It was
first discovered for CO$_{2}$ molecule where the ratio of two molecular
vibration frequencies $\omega _{1}$ and $\omega _{2}$ is close, by accident,
to 1:2. This leads to a quasidegeneracy of the states with energies $\hbar
\omega _{1}$ and  $2\hbar \omega _{2}$. As a result the first harmonic of
the first mode is strongly coupled with the second harmonic of the second
mode which leads to the violation of symmetry-induced selection rules and
appearance of additional lines in the Raman spectra. Fermi resonance is
typical for many organic compounds, especially containing CH group, and is
well investigated by molecular spectroscopists \cite{lisitsa}. The theory of
the Fermi resonance is rather simple since the ultraquantum limit $T\ll
\hbar \bar{\omega}$ (where $\bar{\omega}$ is a mean phonon frequency) is
typical for all these cases and therefore there are only few states which
are involved in the resonance. 

When we consider Fermi-resonance-like phenomena for crystals we immediately
face several difficulties which make the problem much less trivial \cite
{FMM87,JETPLett87,JETP90}. First, the classical limit $T>\hbar \bar{\omega}$
is interesting now and it turns out to be much more complicated than the
ultraquantum one since here we have an infinite number of relevant states.
Second, now we have to satisfy the resonance conditions not only for the
frequencies but also for the phonon wave vectors $\mathbf{q}$. Then, there
is a continuum of nonresonant phonons which is nevertheless also important
working as a thermal bath and leading to the phonon damping and ``Brownian
motion'' fluctuations. The most important point is the occurrence of
essentially new phenomenon in the classical case, namely, a low-frequency
dynamics due to energy transfer between the modes participating in the
resonance; this effect is exponentially small at low temperatures
(ultraquantum regime) when the corresponding modes are not excited. This
``slow'' dynamics corresponds to transitions between the states split by the
anharmonic interaction under the Fermi resonance conditions. In this sense
it relates to the Fermi resonance like the electron paramagnetic resonance
relates to the Zeeman effect. Experimentally it may result in the 
appearance of low-energy peaks in the quasielactic neutron scattering
spectra \cite{FMM87,JETPLett87,JETP90}. 

These new phenomena result from a correlated character of atomic
displacements corresponding to different phonon branches under the Fermi
resonance conditions. In a generic case the smallness of adiabatic parameter
justifies the applicability of the perturbation theory for the consideration
of anharmonic effects \cite{Cowley,KT}. From the technical point of view, it
means the decoupling of the higher-order phonon correlators according to the
Wick's theorem, or, in classical terms, we suppose that the initial phases
of the interacting phonons are random. This is a standard approximation
which is used to calculate the temperature dependence of phonon frequencies
and dampings \cite{VKTdyn,khrom1,khrom2}. However, when the phonon frequency
ratio turns out to be integer (and provided that some symmetry conditions
are satisfied) a well-known nonlinear effect, the ``phase locking'', appears 
\cite{Haken} and the perturbation theory fails. In particular, it leads to
an essential enhancement of multiphonon contributions to the dynamical
structure factor in comparison with the perturbative treatment, and to the
appearance of quasistatic atomic displacements. This is a consequence of
subtle phase coherence phenomena which are completely igonored by a standard
perturbative treatment. The situation is similar to the Anderson
localization in disordered systems which also cannot be described by any
consideration of the \textit{average} Green function \cite{anderson}.

To illustrate general mechanisms consider first the simplest case of the
resonance, namely, phonons with the wave vector $\mathbf{q}_{0}=(\frac{2}{3},%
\frac{2}{3},\frac{2}{3})$ and the frequency ratio 1:2 \cite
{JETPLett87,JETP90}. It corresponds to BCC phases of alkali and alkaline
earth metals (for instance, for potassium the frequencies of longitudinal
and transverse phonons in this point at $T=4.2K$ are 0.27$\pm $0.01 and 0.55$%
\pm $0.01 of ionic plasma frequency, respectively). Transverse phonons are
double degenerate, however, microscopic estimations of the corresponding
anharmonic coupling constants \cite{JETPLett87} show that only one transverse
branch participates in the resonance since the coupling constant for another
one is four orders of magnitude smaller. So we can consider a simplified
model with the interaction of one longitudinal branch (with the displacement
field $u\left( \mathbf{r},t\right) $) and one transverse one (with the
displacement field $v\left( \mathbf{r},t\right) $). Taking into accoun only
resonant interaction term $V=\lambda \int d\mathbf{r} uv^{2}$ one can write a 
set of equations of motion,
\begin{equation}
\left\{ 
\begin{array}{l}
\ddot{u}+\omega ^{2}\left( -i\nabla \right) u+2\gamma \left( -i\nabla
\right) u+\lambda v^{2}=0 \\ 
\ddot{v}+\Omega ^{2}\left( -i\nabla \right) v+2\Gamma \left( -i\nabla
\right) v+2\lambda uv=0
\end{array}
\right.   \label{Newton}
\end{equation}
where $\gamma ,\Gamma $ are corresponding phonon damping parameters. We try
the solutions of Eq.(\ref{Newton}) in the form
\begin{equation}
\left\{ 
\begin{array}{l}
u=A\left( \mathbf{r},t\right) \exp {i}\left[ {\mathbf{q}_{0}\mathbf{r}%
-\omega }\left( {\mathbf{q}_{0}}\right) {t}\right] + \\ 
B\left( \mathbf{r}%
,t\right) \exp {i}\left[ {\mathbf{q}_{0}\mathbf{r}+\omega }\left( {\mathbf{q}%
_{0}}\right) {t}\right] +c.c. \\ 
v=C\left( \mathbf{r},t\right) \exp {i}\left[ {\mathbf{q}_{0}\mathbf{r}%
+\omega }\left( {\mathbf{q}_{0}}\right) {t/2}\right] + \\ 
D\left( \mathbf{r}%
,t\right) \exp {i}\left[ {\mathbf{q}_{0}\mathbf{r}-\omega }\left( {\mathbf{q}%
_{0}}\right) {t/2}\right] +c.c.
\end{array}
\right.   \label{Solut}
\end{equation}
Amplitudes $A,B,C,D$ are slowly varying in space and time functions. It is
important that the wave vector ${\mathbf{q}_{0}}$ is equivalent to $-2{%
\mathbf{q}_{0}}$ since $\exp (3i{\mathbf{q}_{0}\mathbf{r)=}}1$ for the
crystal lattice sites. Substituting Eq.(\ref{Solut}) into Eq.(\ref{Newton}),
taking into account only resonant terms and only leading approximations in
the anharmonic smallness parameter (for more details, see Ref. \cite{JETP90}%
)  
\begin{equation}
\left\{ 
\begin{array}{l}
\frac{\partial A}{\partial t}+\left( \frac{\partial {\omega }}{\partial {%
\mathbf{q}}}\right) _{0}\nabla A- 
\frac{i}{4{\omega }_{0}}\left( \frac{%
\partial ^{2}{\omega }^{2}}{\partial q_{\alpha }\partial q_{\beta }}\right) 
\frac{\partial ^{2}A}{\partial x_{\alpha }\partial x_{\beta }} \\ 
+\gamma_{0}A+i\Lambda C^{\ast 2} =0  \\ 
\frac{\partial C}{\partial t}-\frac{1}{2}\left( \frac{\partial {\omega }}{%
\partial {\mathbf{q}}}\right) _{0}\nabla C+ 
\frac{i}{8{\omega }_{0}}\left( 
\frac{\partial ^{2}{\omega }^{2}}{\partial q_{\alpha }\partial q_{\beta }}%
\right) _{0}\frac{\partial ^{2}C}{\partial x_{\alpha }\partial x_{\beta }} \\ 
+\left( \Gamma _{0}+\frac{i\nu _{0}}{2}\right) C-4i\Lambda A^{\ast }C^{\ast
} =0  
\end{array}
\right. \label{envel}
\end{equation}
where $\omega =2\Omega +\nu $ $\left( \nu <<\omega \right) ,\Lambda =\lambda
/2{\omega }_{0}$ and subscript ``0'' means ${\mathbf{q}=\mathbf{q}_{0}.}$
Equations for $B$ and $D$ differ from Eq.(\ref{envel}) by the replacement $%
\nu \rightarrow -\nu ,\Lambda \rightarrow -\Lambda .$ 

One can demonstrate \cite{JETP90} that at small enough $\nu $ (realistic
estimations for alkali metals show that ``small'' means 5-7\% of $\omega $)
purely sine waves $A,B,C,D=const$ turn out to be unstable with respect to a
self-modulation and the solitons of the envelopes can form. This modulation
is connected with a slow (in comparison with a characteristic phonon times)
dynamics of energy transfer between longitudinal and transverse phonons. 
Numerical simulations of the thermal noise effects \cite{gorn1,gorn2} show
that the latter do not suppress this slow dynamics. It appeared that there
are \textit{two} limit circles in this system (which correspond to two phase
locking regimes with different relative phases for $A$ and $C$ waves) and
this energy transfer dynamics can be described as a stochastic resonance
between these two limit circles \cite{gorn2}. 

Similar phenomena can be also considered for acoustic (long-wavelength)
phonons as an energy transfer between two ultrasound waves with different 
polarization vectors and integer ratio of sound velocities \cite{acoustic}.
Alkali metals near the melting point (1:3 ratio for different transverse
sound waves propagating into <110> direction) or W$_{1-x}$Re$_x$,
Mo$_{1-x}$Re$_x$ alloys (1:1 ratio for the same sound waves) might be
an interesting examples. Computer simulations \cite{acoustic} demonstrate
that, depending on the initial phases of the waves, this energy transfer
may be both chaotic and quasiperiodic. It would be interesting to check
this prediction experimentally.

It is worthwhile to note that for the case of 1:2 frequency ratio considered 
above the resonance conditions for phonon wave vectors can be satisfied only in
some peculiar points of the Brillouin zone. For the case of 1:3 resonance
they can be satisfied in a generic case; for high-symmetry
directions the number of waves participating in the resonance coupling
can be very large. FCC La is an interesting example of such resonance
(below we follow our work \cite{stroev}). The lattice dynamics of FCC La is
characterized by a drastic nonmonotonicity of the dispersion curves $\omega (%
\mathbf{q})$ for transverse phonons in $\mathbf{q}\parallel \langle
111\rangle $ direction and by a significant temperature dependence of their
frequencies, i.e. by strong anharmonic interactions \cite{l6}. As it was
discussed in Ref. \cite{l6} this behavior can be accounted for by the
electron-phonon interaction (Kohn anomalies). One can see from these
experimental data that the ratio of longitudinal to transverse phonon
frequencies in $\langle 111\rangle $ direction varies around the integer
ratio $\frac{\omega _{2}(\mathbf{q}_{0})}{\omega _{1}(\mathbf{q}_{0})}=3$,
the value of wave vector $\mathbf{q}_{0}$ varying with the temperature
within sufficiently wide limits. This makes it possible to shift the
resonance watching simultaneously the change in the lattice dynamics
behavior.

Let us set up the simplest model to describe the resonance effects under
consideration in the lattice dynamics of FCC La. We shall proceed with the
equations of motion for the amplitudes of longitudinal $(u)$ and transverse $%
(v)$ phonons allowing only for the ``resonance'' anharmonic interaction $%
V=\lambda \int d\mathbf{r}uv^{3}$:

\begin{equation}
\left\{ 
\begin{array}{l}
\ddot{u}+\omega ^{2}u+\lambda v^{3}=0 \\ 
\ddot{v}+\Omega ^{2}v+3\lambda uv^{2}=0
\end{array}
\right.   \label{Newton1}
\end{equation}

Because of the FCC lattice symmetry we should consider phonons propagating
in four equivalent directions $\langle 111\rangle ,\langle 1\bar{1}\bar{1}%
\rangle ,\langle \bar{1}1\bar{1}\rangle ,\langle \bar{1}\bar{1}1\rangle $;
the relevant wave vectors meeting condition $\omega (\mathbf{q})=3\Omega (%
\mathbf{q})$ will be denoted as $\mathbf{q}_{j}$ ($j=0,1,2,3$). Similarly to
Ref. \cite{JETP90} we try the solutions of the equations (\ref{Newton1})
as 
\begin{equation}
\left\{ 
\begin{array}{l}
u=\sum\limits_{j=0}^{3}(A_{j}\exp {i(\mathbf{q}_{j}\mathbf{r}-\omega t)}%
+B_{j}\exp {i(\mathbf{q}_{j}\mathbf{r}+\omega t)})+c.c. \\ 
v=\sum\limits_{j=0}^{3}(C_{j}\exp {i(\mathbf{q}_{j}\mathbf{r}-\Omega t)}%
+D_{j}\exp {i(\mathbf{q}_{j}\mathbf{r}+\Omega t)})+c.c.
\end{array}
\right.   \label{phonons}
\end{equation}
where $A_{j},B_{j},C_{j},D_{j}$ are slowly varying (due to $\lambda $
smallness) functions of $\mathbf{r}$ and $t$, i.e. the envelopes of the
phonons under consideration. Substituting Eqs. (\ref{phonons}) into Eqs. (%
\ref{Newton1}) and neglecting the second time derivatives of the envelopes
as well as the nonresonance terms we obtain the following set of equations 
\begin{equation}
\left\{ 
\begin{array}{l}
\dot{A_{0}}+i\Lambda (6D_{1}^{\ast }D_{2}^{\ast }D_{3}^{\ast
}+3C_{0}^{2}D_{0}^{\ast }+6C_{0}\sum\limits_{j=1}^{3}C_{j}D_{j}^{\ast })=0
\\ 
\dot{B_{0}}-i\Lambda (6C_{1}^{\ast }C_{2}^{\ast }C_{3}^{\ast
}+3D_{0}^{2}C_{0}^{\ast }+6D_{0}\sum\limits_{j=1}^{3}D_{j}C_{j}^{\ast })=0
\\ 
\dot{C_{0}}+9i\Lambda (2B_{1}^{\ast }C_{2}^{\ast }C_{3}^{\ast }+2B_{2}^{\ast
}C_{1}^{\ast }C_{3}^{\ast }+2B_{3}^{\ast }C_{1}^{\ast }C_{2}^{\ast }+ \\ 
\qquad \qquad \qquad 2A_{0}\sum\limits_{j=0}^{3}C_{j}^{\ast
}D_{j}+2D_{0}\sum\limits_{j=1}^{3}A_{j}C_{j}^{\ast }+ \\ 
\qquad \qquad \qquad \qquad D_{0}^{2}B_{0}^{\ast
}+2D_{0}\sum\limits_{j=1}^{3}D_{j}B_{j}^{\ast })=0 \\ 
\dot{D_{0}}-9i\Lambda (2A_{1}^{\ast }D_{2}^{\ast }D_{3}^{\ast }+2A_{2}^{\ast
}D_{1}^{\ast }D_{3}^{\ast }+2A_{3}^{\ast }D_{1}^{\ast }D_{2}^{\ast }+ \\ 
\qquad \qquad \qquad 2B_{0}\sum\limits_{j=0}^{3}D_{j}^{\ast
}C_{j}+2C_{0}\sum\limits_{j=1}^{3}B_{j}D_{j}^{\ast }+ \\ 
\qquad \qquad \qquad \qquad C_{0}^{2}A_{0}^{\ast
}+2C_{0}\sum\limits_{j=1}^{3}C_{j}A_{j}^{\ast })=0;
\end{array}
\right.   \label{envelopes}
\end{equation}
the rest twelve equations are obtained from (\ref{envelopes}) by cyclic
permutations of the indices. Here $\Lambda =\frac{\lambda }{2\omega }$.

The dynamics of the energy transfer in this system has been investigated in
Ref. \cite{stroev} by numerical simulations. It was shown that depending on
the initial phases of the involved phonons this energy transfer can be
either regular or chaotic. One can expect from the dimension considerations
that a typical frequency of this slow dynamics will be of order of 
\begin{equation}
\omega ^{\ast }\simeq \bar{\omega}(\frac{\bar{u}^{2}}{a^{2}})\frac{\lambda }{%
M\bar{\omega}^{2}a^{6}},  \label{unit_freq}
\end{equation}
Here $M$ is the ion mass, $a$ is the lattice constant, $\bar{u}^{2}$ is the
average square of atomic displacements. For the room temperature $\omega
^{\ast }\approx 10^{-3}\bar{\omega}$ can be taken for the estimation.
However, it was shown that in reality it is much larger, $\triangle \omega
\simeq 10^{2}\omega ^{\ast },$ due to a large number of interacting waves
(32 real fields). 

Similarly to Ref. \cite{JETP90} (see also above) it can be shown that the
account for intermode or non-resonant intramode anharmonicities may result
in the appearance of the peaks in quasielastic neutron scattering spectra
with a frequency width about $\triangle \omega $. The largest contributions
result from the lowest-order (three-phonon) anharmonic processes. For
example, the term $V=\mu \int d\mathbf{r}v^{3}$ in the potential energy
leads to the appearance of the ``low-frequency'' contribution to the
transverse phonon field, $\delta v=-3\mu v^{2}/\Omega ^{2}$. Assuming all
the amplitudes $A_{j}$ and $C_{j}$ in the equation (\ref{phonons}) to be
constant we would obtain static contributions to $v$ with the wave vectors $%
\mathbf{Q}_{a}=\mathbf{q}_{i}\pm \mathbf{q}_{j}+\mathbf{g}$ where $\mathbf{g}
$ is a reciprocal lattice vector (which can be, in particular, equal to
zero). One can expect therefore the peaks in the quasielastic neutron
scattering spectra near the wavevectors $\mathbf{Q}_{a}$ which reflect the
dynamics of the envelopes. In a special case considered above for the
frequency ratio 2:1 one of these vectors coincides with the vector $\mathbf{q%
}_{i}$ but for the ratio 3:1 it is impossible. The vectors $\mathbf{Q}_{a}$
for $i=j$ are just the reciprocal lattice vectors but the rest of them
should be temperature dependent due to the temperature dependence of the
phonon frequencies shifting the Fermi resonance point. It can be important
for the experimental verification of the effects under consideration. A
similar contribution of order of $u^{2}$ appears in the longitudinal phonon
field because of the potential energy term $V=\nu \int d\mathbf{r}u^{3}$.

\section{Conclusions}

To conclude, we predict a new class of nonperturbative anharmonic phenomena
in lattice dynamics under resonance conditions in the phonon system. This resonance
results in the instability of phonons and complicated picture of energy transfer
between the modes in the resonance. It can be experimentally investigated by
inelastic and quasielastic neutron scattering as well as by acoustic methods.
There is an interesting open issue what is the role of these phenomena
in the development of lattice instabilities of the crystals, in particular,
in the melting \cite{FMM87}. This question is motivated by two general remarks:
(i) anharmonic effects in a generic case are small up to the melting point
and (ii) instability of a generic weakly anharmonic dynamical system, according
to the KAM theory, is connected with the resonances. Therefore it might appear that
the phenomenon under consideration is not a kind of exotic but important
for any crystal.

\end{document}